\documentclass[a4paper,11pt]{article}
\usepackage{pos}
\newcommand{\dph}{$\Delta\varphi \ $}

\title{Angular correlations of heavy-flavour hadron decay electrons and charged particles in pp collisions at $\sqrt{s}$ =
5.02 TeV with ALICE at the LHC}
\ShortTitle{Heavy flavor electron correlation with ALICE}
\manuallySeparateAuthors

\author*[a]{Ravindra Singh}
\author{for the ALICE Collaboration}

\affiliation[a]{Department of Physics, Indian Institute of Technology Indore,\\
  Indore, INDIA}

\emailAdd{ravindra.singh@cern.ch}

\abstract{Two-particle azimuthal correlations of electrons from heavy-flavour hadron decays with charged particles can give insight into the properties of heavy-quark production and hadronization into heavy-flavour jets.\\
In this contribution, the ALICE results measured in pp collisions at 5.02 TeV, collected in the LHC Run 2,  are presented. The yields of charged particles in the near and away-side correlation peaks and the peak width are compared with PYTHIA calculations.}

\FullConference{%
 
 The Ninth Annual Conference on Large Hadron Collider Physics - LHCP2021\\
 7-12 June 2021\\
 Online
}

\begin{document}
\maketitle

\section{Introduction}

Heavy flavours (charm and beauty quarks), due to their large masses, mainly originate in the initial hard scatterings of partons from the incoming
beam particles, evolve as parton
showers and hadronize. They are finally observed as back-to-back
jet events~\cite{Cunqueiro:2015dmx}. Jet-like correlation studies can be used to obtain information on parton energy loss and modification of their fragmentation function~\cite{ALICE:2016gso} in the medium formed in high-energy heavy-ion collisions (AA). The study of azimuthal correlations in pp collisions is useful as a reference to investigate the properties of the medium, while measurements performed in p-A collisions are important to study possible modifications of the fragmentation functions due to cold nuclear matter effect or gluon saturation effects. Furthermore, in pp collisions, heavy-flavour correlation measurements can be used to test the predictions of pQCD calculations as the heavy-flavour masses are larger than the perturbative QCD scale ($m_{\rm{c,b}} >> \Lambda_{\rm{QCD}}$).

\begin{figure}[ht]
\begin{center}
\includegraphics[scale=.34]{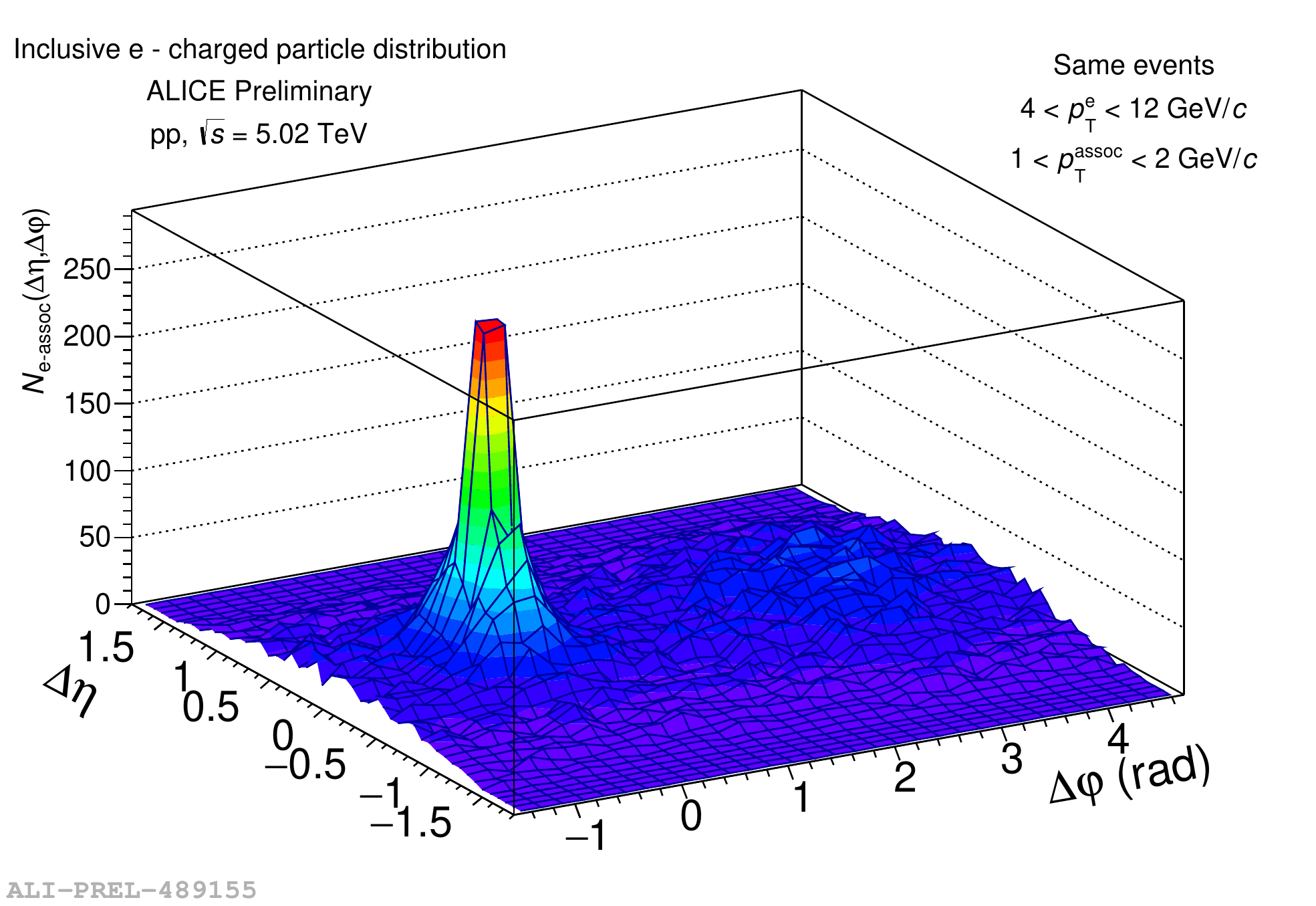}
\includegraphics[scale=.32]{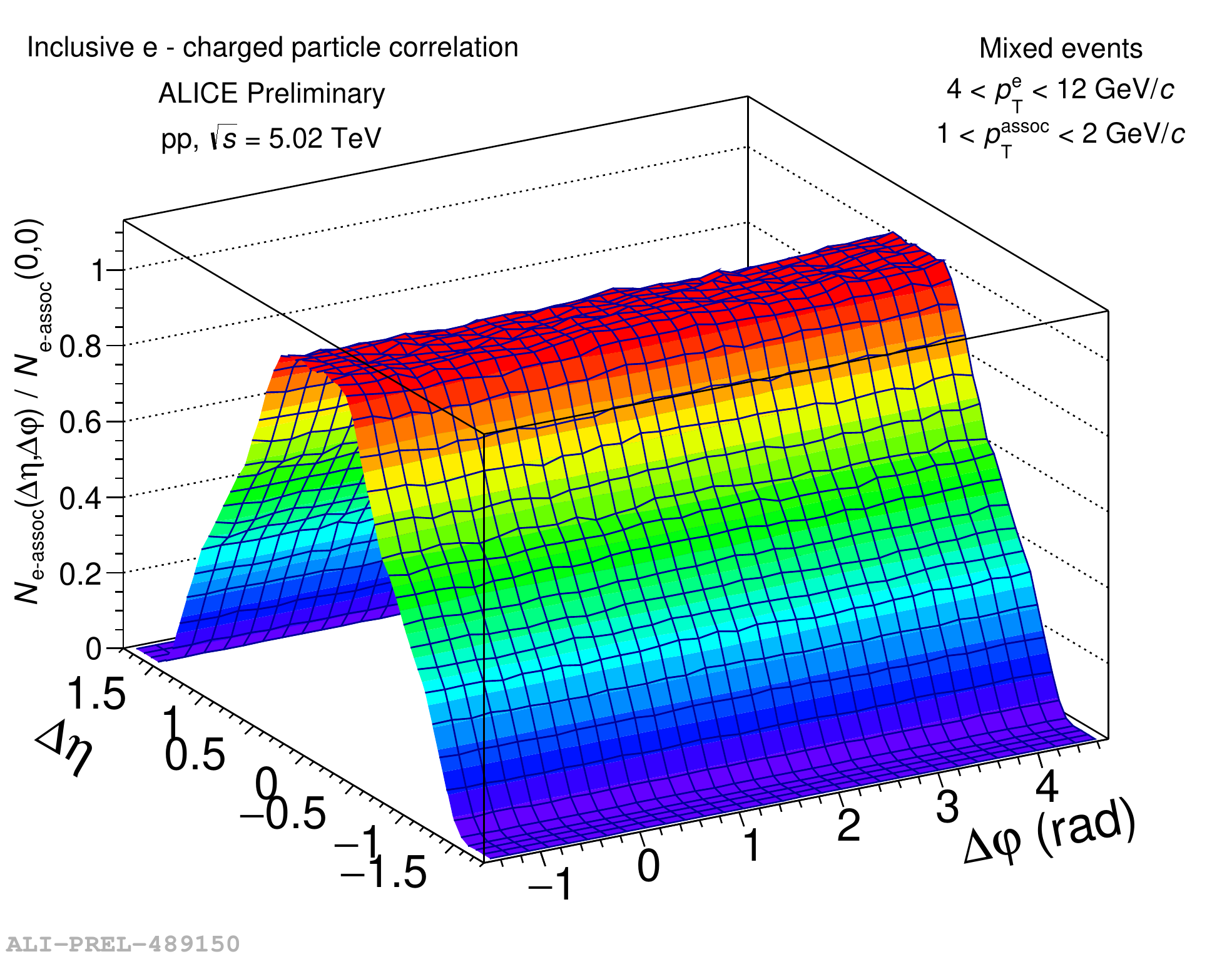}
\includegraphics[scale=.34]{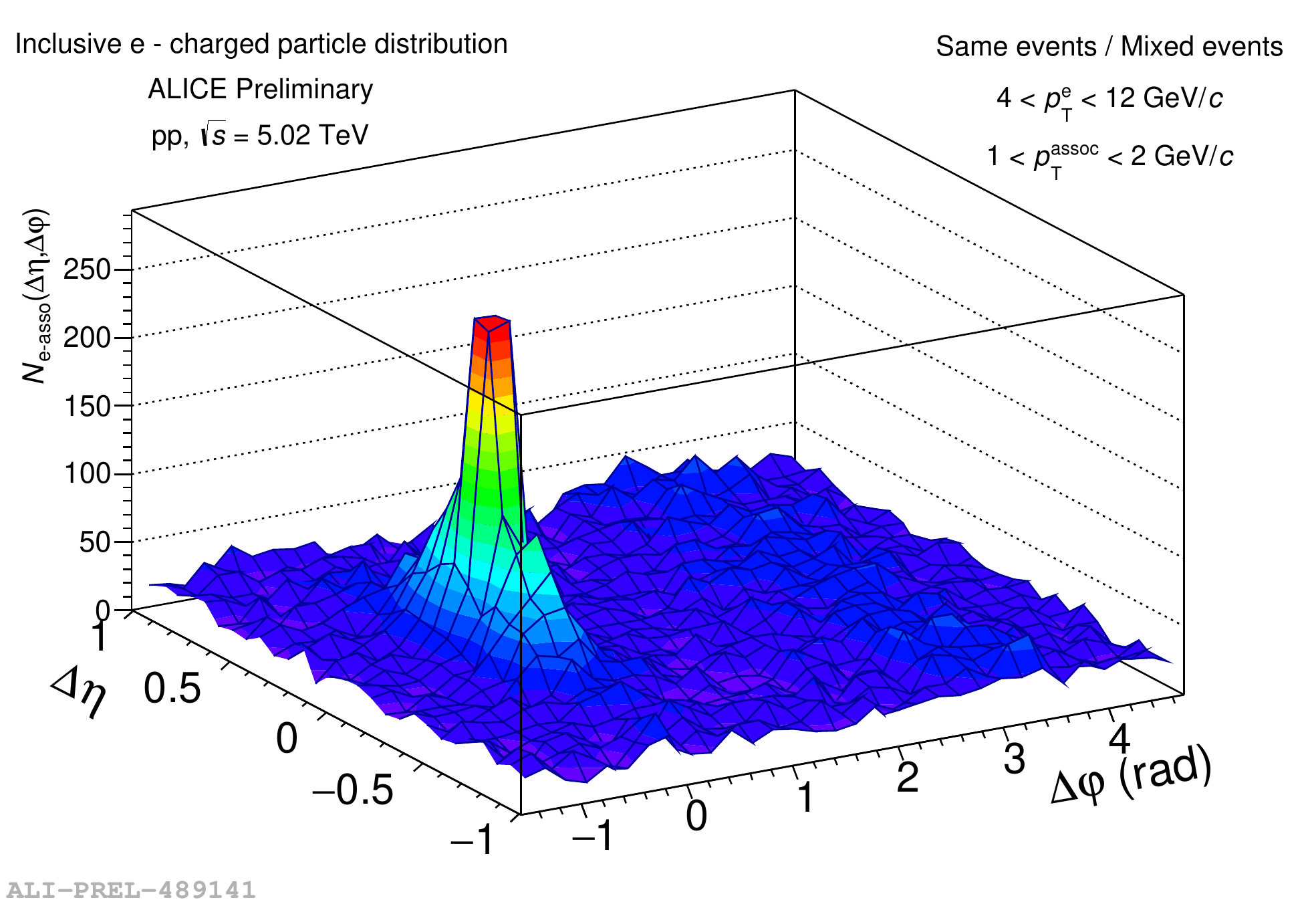}
\caption{The $\bigtriangleup\varphi$ and $\bigtriangleup\eta$ distribution between inclusive electrons and charged hadrons for $1 < p_{\rm T}^{h} < 2$ GeV/$c$, same-event (upper left), mixed-event (upper 
right), and same-event/mixed-event (bottom). }
\label{fig:fig3}
\end{center}
\end{figure}

\section{Analysis Strategy}

Electrons from heavy flavour hadron decays are identified using the Time Projection Chamber (TPC) and Electromagnetic Calorimeter (EMCal) in ALICE~\cite{alice}. Particle identification in the TPC is performed by measuring the specific ionization energy loss in the detector gas. The EMCal detector identifies electrons by their $E$ (energy) over $p$ (momentum) distribution with a shower shape selection~\cite{ALICE:2019bfx}. The background is composed mainly by non-heavy-flavor (Non-HFE) electrons originating from $\gamma$ conversions and Dalitz decays; they are identified by studying the invariant mass distributions of electron pairs. Those which have invariant mass less than 140 MeV/$c^{2}$, are tagged as Non-HFE. The non-HFE finding efficiency (tagging efficiency) is $\sim 75\%$, estimated using Monte Carlo (MC) simulations. The remaining
Non-HFE contamination in the heavy flavour electron (HFE) sample is corrected by using the tagging
efficiency.\\

\begin{figure}[ht]
\begin{center}
\includegraphics[scale=0.6]{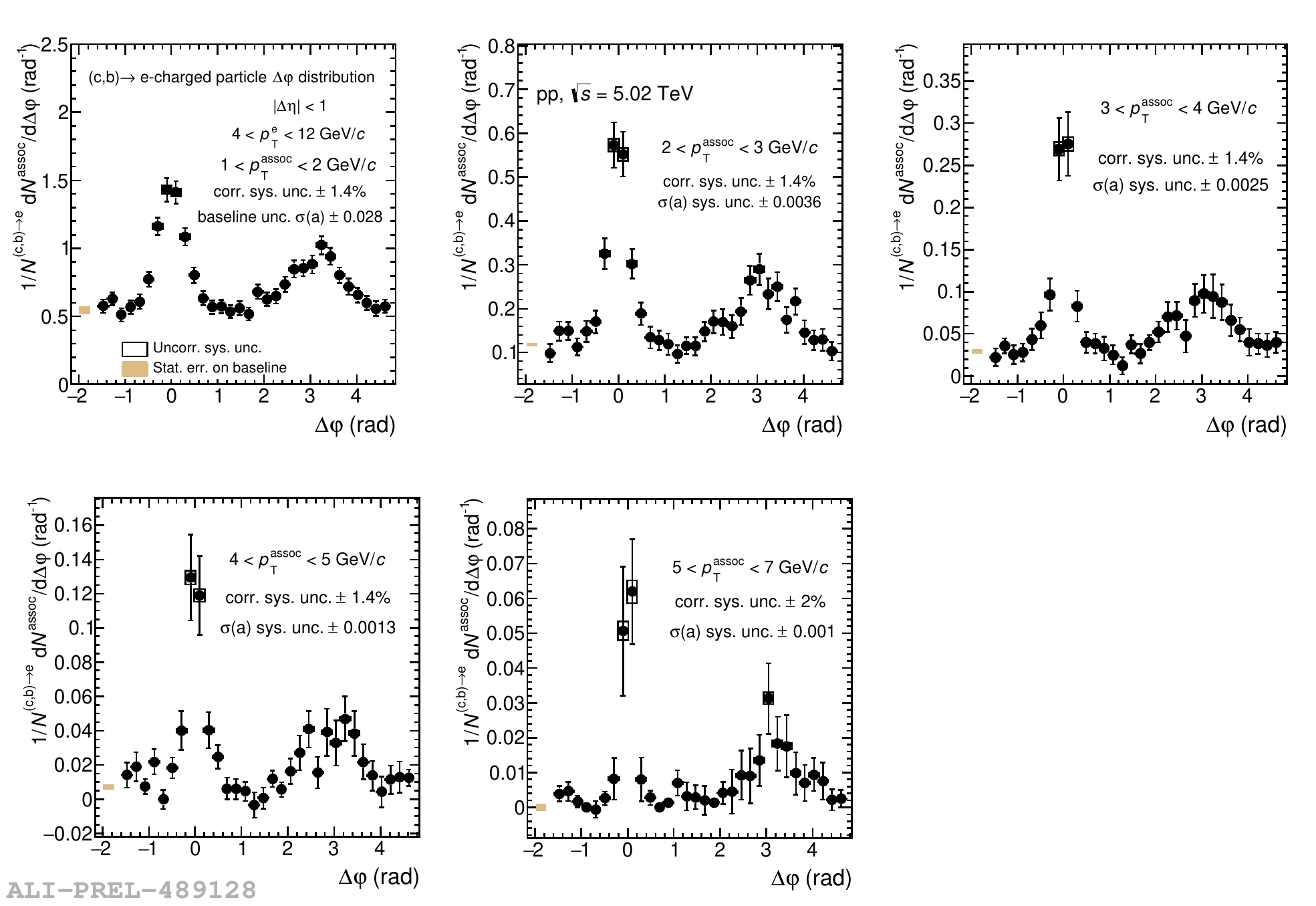}
\caption{The $\bigtriangleup\varphi$ distribution between heavy-flavour decay electrons and charged hadrons, normalized by the total number of triggered heavy-flavour decay electrons are shown for five different associated particles $p_{\rm T}$ intervals, from $1 < p_{\rm T} < 7$ GeV/$c$.}
\label{fig:delphinorm}
\end{center}
\end{figure}

\begin{figure}[ht]
\begin{center}
\includegraphics[scale=0.6]{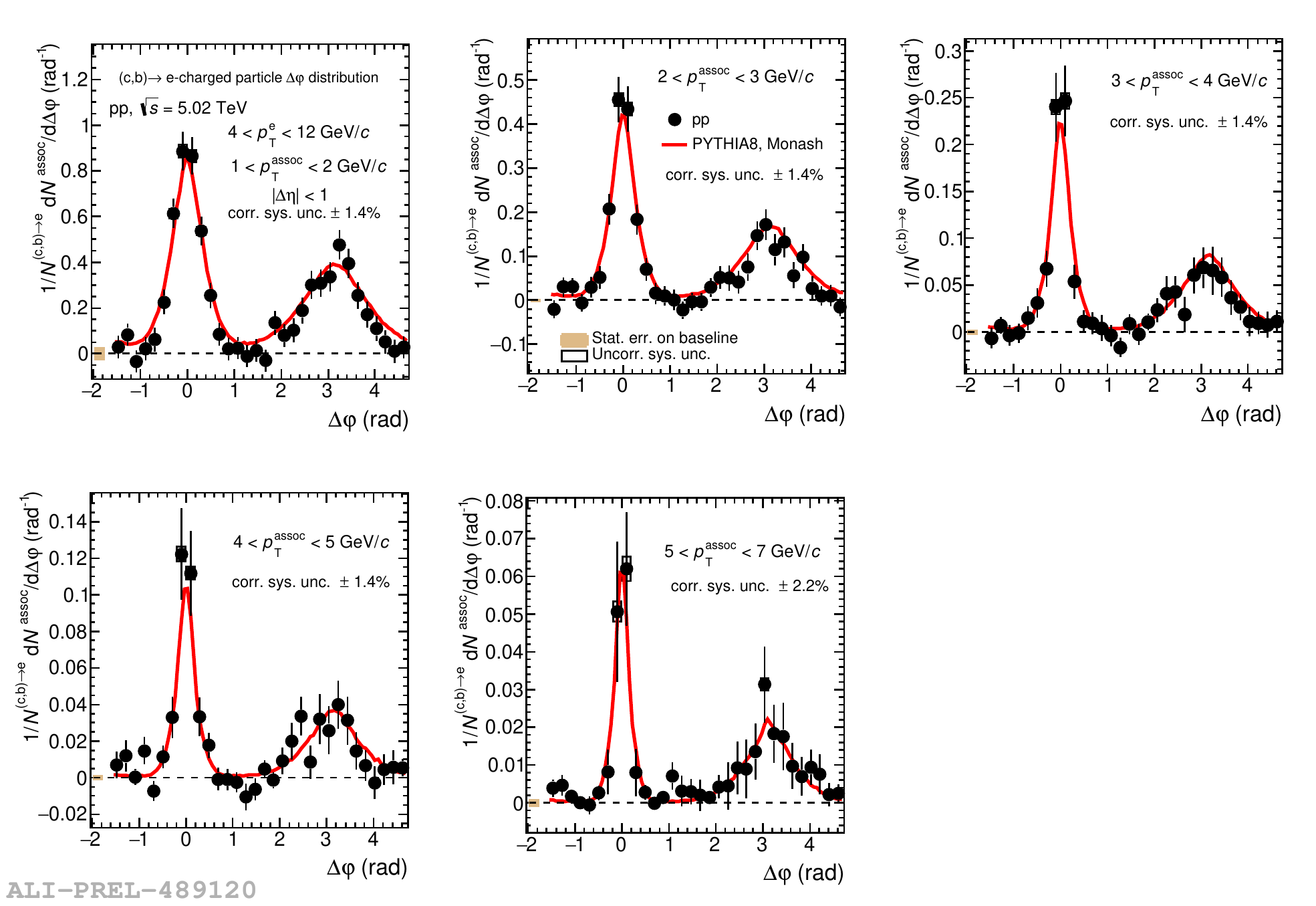}
\caption{The baseline subtracted $\bigtriangleup\varphi$ distribution between heavy-flavour decay electrons and charged hadrons are shown for five different associated particles  $p_{\rm T}$ intervals, from $1 < p_{\rm T} < 7$ GeV/$c$.}
\label{fig:delphiHFE}
\end{center}
\end{figure}

\begin{figure}[ht!]
\begin{center}
\includegraphics[scale=.33]{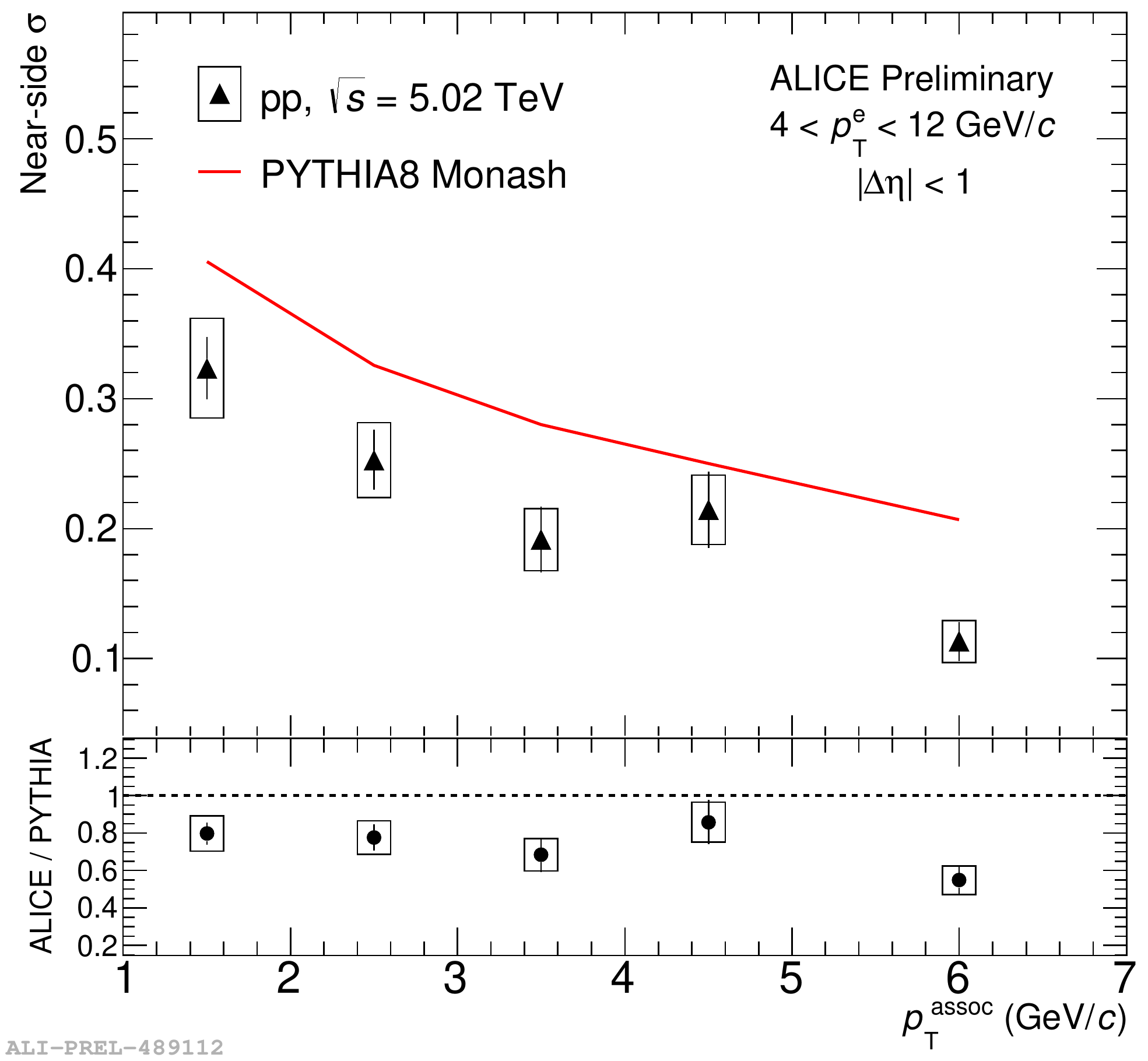}
\includegraphics[scale=.365]{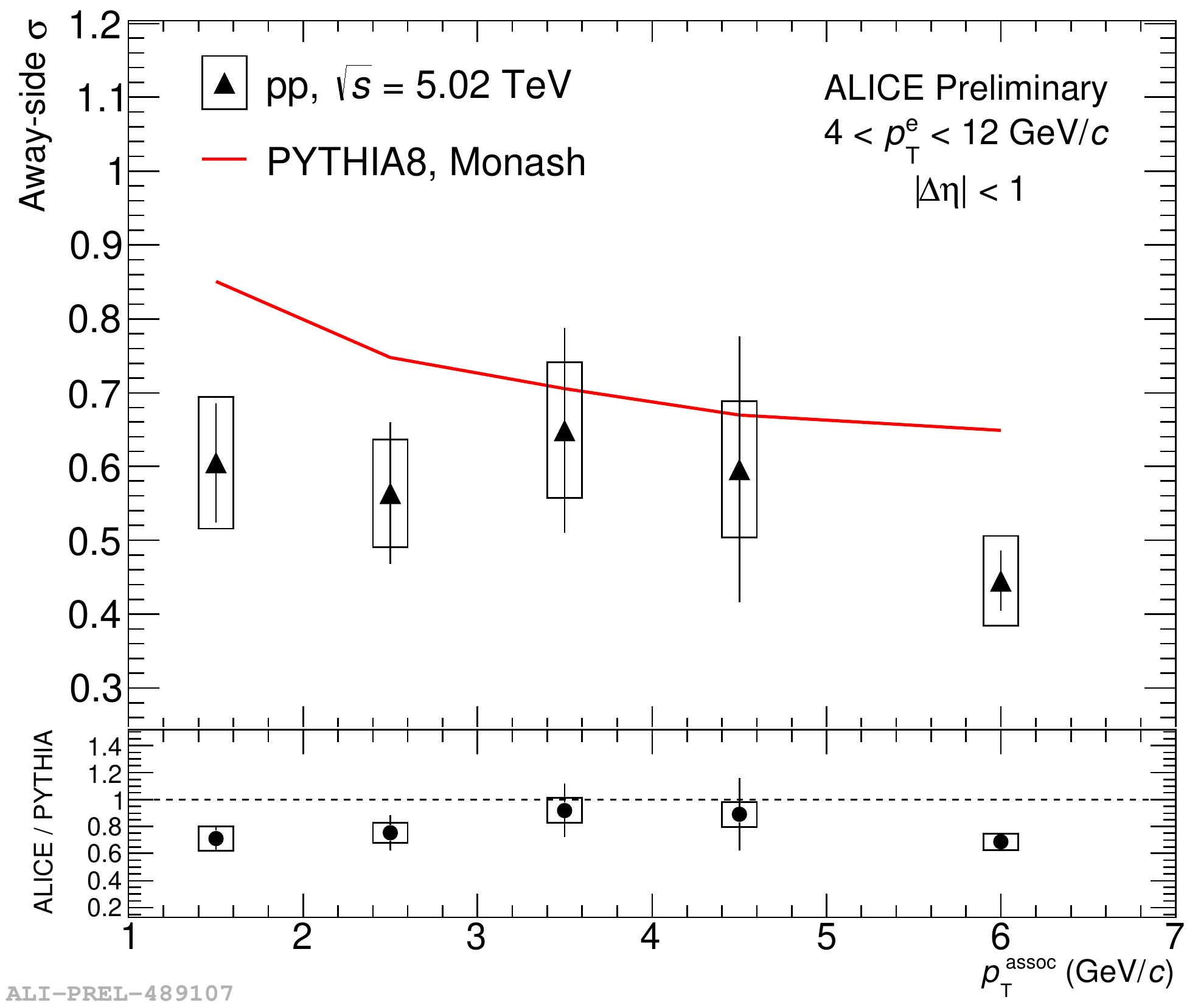}
\caption{Near-side (left) and away-side (right) width ($\sigma$) compared with PYTHIA8 (Monash).}
\label{fig:figwidth}
\end{center}
\end{figure}

\begin{figure}[ht!]
\begin{center}
\includegraphics[scale=.353]{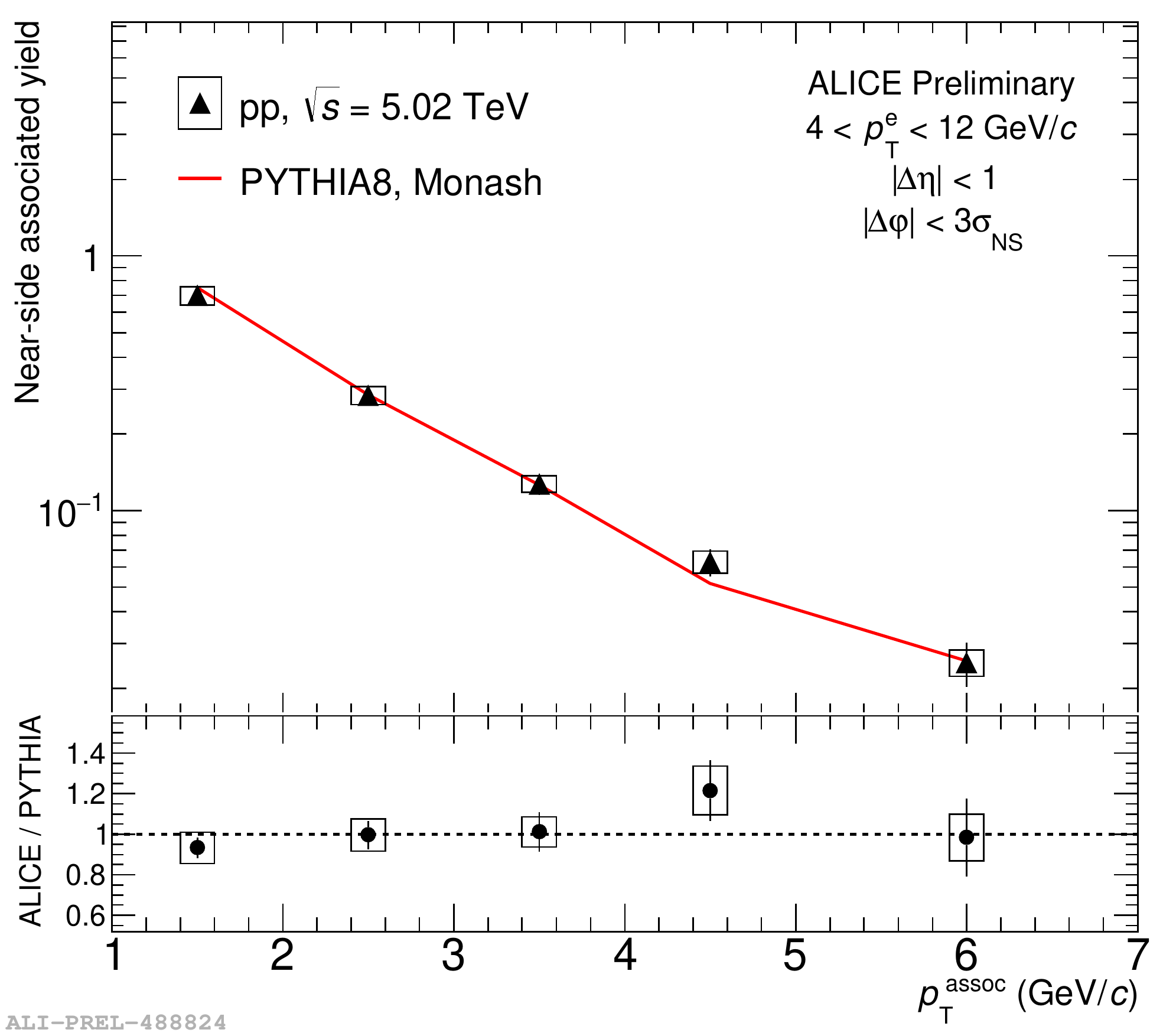}
\includegraphics[scale=.347]{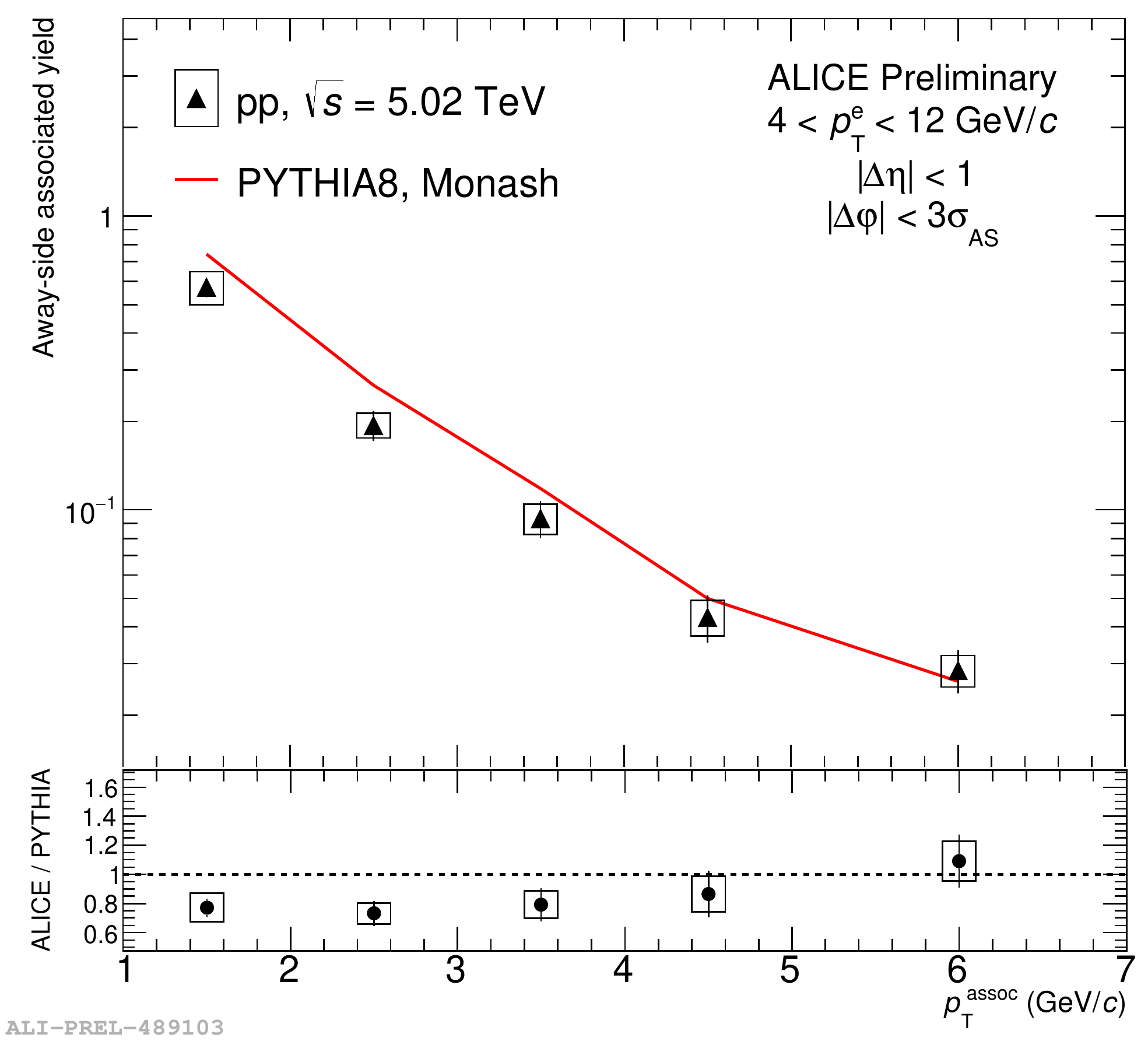}
\caption{Near-side (left) and away-side (right) associated yield compared with PYTHIA8 (Monash).}
\label{fig:figyield}
\end{center}
\end{figure}

The azimuthal angular correlation between heavy-flavour decay electrons and charged hadrons is obtained after detector efficiency and acceptance corrections. These corrections are performed via the mixed-event technique~\cite{ALICE:2016gso}, in which the \dph distribution from the correlation of trigger and associate particles of the same event is divided by the corresponding distribution among mixed events, as shown in Figure~\ref{fig:fig3}. 

The contamination from hadrons is accounted for by subtracting a di-hadron azimuthal correlation from the obtained inclusive electron-hadron correlation. The Non-HFE \dph distribution is obtained by subtracting the \dph distributions of the like-sign from unlike-sign electrons, which is further corrected by the tagging efficiency. The HFE-hadron correlation function is determined by subtracting the Non-HFE \dph distribution from the inclusive electron \dph distribution.

\section{Results}
In Figure~\ref{fig:delphinorm}, the $\bigtriangleup\varphi$ distributions between heavy-flavour decay electrons ($4< p_{\rm T}^{e} < 12$ GeV/$c$) and charged hadrons ($1 < p_{\rm T}^{assoc} < 7$ GeV/$c$) are normalized by the total number of triggered heavy-flavour decay electrons. In this distribution,  the near-side peak at \dph = 0 is formed by the particles associated with the trigger particle, whereas the away-side peak at \dph = $\pi$ originates from back-to-back dijets.

The pedestal (baseline) is subtracted from the distribution using a generalized Gaussian fitting function:\\

\begin{footnotesize}
\begin{equation}
{f(\Delta\varphi)} = b + \frac{Y_{NS}\times\beta_{NS}}{2\alpha_{NS}\Gamma{(1/\beta_{NS})}}\times e^{-(\frac{\Delta\varphi}{\alpha_{NS}})^{\beta_{NS}}} + \frac{Y_{AS}\times\beta_{AS}}{2\alpha_{AS}\Gamma{(1/\beta_{AS})}}\times e^{-(\frac{\Delta\varphi - \pi}{\alpha_{AS}})^{\beta_{AS}}},
\label{eq:fitfunggaus}
\end{equation}
\end{footnotesize}

Here, $b$ is the the baseline, which is a flat contribution of yield below the peaks, NS and AS refer to near-side and away-side observables respectively. $Y$ is the associate particle yield. The $\alpha$ parameter is related to the variance, and the $\beta$ parameter characterizes the shape of peaks. The distribution becomes a Gaussian for $\beta =2$. The widths of the peaks are obtained using the relation~\cite{ALICE:2019oyn}:

\begin{equation}
 \sigma =    \alpha \sqrt{\Gamma{(3/\beta)}/\Gamma{(1/\beta)}}
 \label{eq:sigmarel}
\end{equation}

The baseline subtracted $\bigtriangleup\varphi$ distribution is  compared with PYTHIA8 (Monash) calculations~\cite{Skands:2014pea} as shown in Figure~\ref{fig:delphiHFE}. The PYTHIA8 predictions are consistent with the results. The widths ($\sigma$) of near and away-side peaks are obtained using the Eq.~\ref{eq:sigmarel} and yields of near and away-side peaks are obtained by a bin counting method within |\dph| < $3\sigma_{\rm{NS}}$ and |\dph - $\pi$| < $3\sigma_{\rm{AS}}$ respectively. The near and away-side widths and yields are compared with PYTHIA8 shown in Figures~\ref{fig:figwidth} and~\ref{fig:figyield}, respectively. It is observed that near and away-side widths are consistent with PYTHIA8 within $1-2\sigma$. The near-side yield is consistent over the entire $p_{\rm T}$ range with PYTHIA8, however at low $p_{\rm T}$, the away-side yield in PYTHIA8 overestimates the data by $\sim 20\%$, whereas at high $p_{\rm T}$, it is consistent. \\

\section{Summary}

In this study, azimuthal angular correlations between heavy-flavour decay electrons and charged hadrons in pp collisions at $\sqrt{s} = $ 5.02 TeV are discussed. The \dph distributions, near and away side width and yields for electrons from heavy flavour are measured, and compared with PYTHIA8 predictions. The predictions are in agreement with data, within uncertainties, suggesting a good description of the fragmentation processes in PYTHIA8. .

\pagebreak

\end{document}